\documentclass[12pt,3p,authoryear]{elsarticle}
\usepackage[T1]{fontenc}
\usepackage[utf8]{inputenc}
\usepackage{amsmath,amssymb,amsfonts}
\usepackage{natbib}
\usepackage{graphicx}
\usepackage{epstopdf}
\usepackage{url}

\usepackage[hidelinks]{hyperref}
\usepackage{booktabs}
\usepackage{rotating}
\usepackage{times}
\usepackage[table]{xcolor}
\usepackage{bm}
\usepackage{xspace}
\usepackage{dcolumn}
\newcolumntype{.}[1]{D{.}{.}{#1}}
\newcolumntype{,}[1]{D{,}{,}{#1}}
\usepackage{setspace}

\usepackage{color}
\usepackage{subdepth}
\usepackage{framed}
\usepackage[table]{xcolor}

\usepackage{float}
\usepackage[font={small,singlespacing},labelfont={bf,small},skip=0.2cm]{caption}
\floatstyle{plaintop}
\restylefloat{table}

\restylefloat{figure}			
\journal{Economics Letteres}
\graphicspath{{./img/}}
\usepackage{multirow}
\usepackage{caption}
\RequirePackage{ae,fancyvrb}

\usepackage{upquote}
\DefineVerbatimEnvironment{CodeInput}{Verbatim}{fontshape=sl}
\DeclareMathAlphabet\mathbfcal{OMS}{cmsy}{b}{n}
\usepackage{subfig}

\usepackage{color}
\newcommand{\ie}{\emph{i.e.}\xspace}
\newcommand{\eg}{\emph{e.g.}\xspace}

\newcommand{\insertfloat}[1]{%
\begin{center}
[Insert~#1 about here.]%
\end{center}%
}

\begin{document}
\begin{frontmatter}
\title{A Century of Economic Policy Uncertainty Through\\the French-Canadian Lens\tnoteref{label1}}
\tnotetext[label1]{We are grateful to the Biblioth\`{e}que et Archives Nationales du Qu\'{e}bec via the CO.SHS Corpus BAnQ initiative (\url{https://co-shs.ca/en}), the Canada Foundation for Innovation (\url{https://www.innovation.ca}), and CBC/Radio-Canada (\url{https://cbc.radio-canada.ca}) for providing us with the data and to IVADO and the Swiss National Science Foundation (grants \#179281 and \#191730) for their financial support. A special thanks to
Stéphane Caron and Dany Plourde for making this project possible. We also thank the Editor (Eric Young), an anonymous referee, Andres Algaba, Noemi Ardia, Scott R. Baker, Clément Aymard, Kris Boudt and Carlos Ordas for their comments and
Andres Algaba, Samuel Borms, Kris Boudt and Jeroen van Pelt for providing us with their French EPU lexicon. We acknowledge Noemi Ardia and Michel Veilleux for their unconditional support.}
\author[hec]{David Ardia\corref{cor1}}
\ead{david.ardia@hec.ca}
\cortext[cor1]{Corresponding author. HEC Montréal, 3000 Chemin de la Côte-Sainte-Catherine, Montreal, QC H3T 2A7. Phone: +1 514 340 6103.}
\author[hec,ghent]{Keven Bluteau}
\ead{keven.bluteau@hec.ca}
\author[hec]{Alaa Kassem}
\ead{alaa.kassem@hec.ca}
\address[hec]{Department of Decision Sciences, HEC Montréal, Montréal, Canada}
\address[ghent]{Department of Economics, Ghent University, Belgium\\[1cm]
\large Economics Letters, 2021, volume 205, 109938\\
\url{https://doi.org/10.1016/j.econlet.2021.109938}}

\begin{abstract} 
A novel token-distance-based triple approach is proposed for identifying EPU mentions in textual documents.  The method is applied to a corpus of French-language news to construct a century-long historical EPU index for the Canadian province of Quebec. 
The relevance of the index is shown in a macroeconomic nowcasting experiment.
\end{abstract}
\begin{keyword}
Economic uncertainty \sep Policy uncertainty \sep Canada \sep Quebec \sep token-distance-based triple \sep issues \sep nowcasting \sep Sentometrics 
\JEL C43 \sep C53 \sep D8 \sep E60 
\end{keyword}
\end{frontmatter}

\onehalfspacing

\section{Introduction}

\noindent
The Economic Policy Uncertainty (EPU) index developed by \citet{BakerEtAl2016} is an important indicator in economic and financial applications \citep[see, \eg,][]{CaldaraEtAl2019,AliagaDiazEtAl2019,BrogaardDetzel2012,ChenEtAl2016,Sum2012}. Following the seminal American EPU index, several country/region-specific indices have been developed \citep[see, \eg,][]{AlgabaEtAl2020,GhirelliEtAl2019,DonadelliEtAl2020}, including a Canadian index that uses articles from five English-language Canadian newspapers and starts in 1985. This note uses a unique French Canadian news corpus to construct more than a century of Canadian EPU from the province of Quebec's perspective. Quebec is a key player in the Canadian economy; it has more than 20\% of the Canadian population and is the second contributor to Canadian GDP after Ontario. About 80\% of Quebec residents use French as their main language.

Our data set and the historical aspect of our index raise some challenges. First, the archives of several news sources are grouped by issue rather than by individual articles, which renders the standard EPU count impossible. Second, the availability of archives for particular media sources differs over time, making the static-window standardization used in the traditional sources' aggregation biased.

We propose a novel token-distance-based triple approach to identify EPU mentions in textual documents and use a dynamic normalization for the sources' aggregation to address these problems. We believe our methodology will help researchers compute EPU-like measures from a wider range of textual sources (both in terms of variety and availability) in future studies. We show that our EPU index spikes at major regional and worldwide events. Finally, we show that our index outperforms the existing Canadian EPU when nowcasting major Canadian and Quebec macroeconomic variables. 

\section{Data and methodology}

\subsection{Data}

\noindent 
We consider four sources of French-language news available in Quebec. Archives for three newspapers are retrieved from the Biblioth\`{e}que et Archives Nationales du Qu\'{e}bec (BAnQ): (i) La Presse (January 1928 - December 2013), (ii) Le Devoir (January 1910 - December 2011) and (iii) Le Soleil (January 1972 - December 2006). The original daily issues for these three sources are converted into a computer-readable format using optical character recognition (OCR) technology. The fourth source consists of French-language news articles from Radio-Canada (the francophone arm of the Canadian Broadcasting Corporation) from January 2003 to June 2020. We process the raw data set the standard way. We transform our textual documents (issues and articles) into vectors of tokens using the Word Boundary Rules defined in the Unicode\textsuperscript{\textcopyright} Standard.\footnote{Tokens are the individual units of texts; see \url{http://www.unicode.org/reports/tr29/\#Word\_Boundaries}. We remove any tokens that are French stop words (\eg, ``le''), punctuation, and that contain numeric or non-alphanumeric characters.}\footnote{To illustrate our text processing, the sentence ``Le salaire minimum passera à 13,50\$ l’heure samedi.'', is represented by the five-token vector [salaire, minimum, passera, heure, samedi] at the end of the process. More details is available in the Internet appendix.}


\subsection{Index construction methodology}

\noindent 
Our index construction deviates from \citet{BakerEtAl2016} in three aspects: (i) dictionary, (iii) EPU count and (iii) source aggregation.

\paragraph{Dictionary} 

As we are examining French documents, we rely on a list of French EPU keywords instead of the original list of English keywords. We use the dictionary provided by \citet{AlgabaEtAl2020}, which is based on a translation-enhanced word2vec approach, and add a few tax-related words specific to Canada or Quebec.\footnote{We also test several enhancements of the dictionary using Latent Dirichlet Allocation on the news selected at EPU spikes over various periods. Results with alternative lexicons are qualitatively similar.} 

\paragraph{EPU count} 

For three of our sources, news is aggregated by issue. It is impossible to identify individual articles in an automated way, so we cannot perform the standard article-based EPU count. Instead, we try to identify EPU mentions by measuring the distance between tokens in EPU triples contained in each document. First, we determine the token positions of all EPU keywords occurring in the text. From these identified keywords, we construct all possible EPU triples for which we measure the maximum relative distance between the token-triple positions (\ie, the maximum of the absolute difference between the token position of the E-P, E-U and P-U keywords).\footnote{If a keyword comprises more than one token, leading to multiple token positions for the keyword and thus multiple possible distances for a given triple, the minimum distance is taken.}  We only keep triples with a maximum distance less than or equal to a given threshold~$\tau$. The threshold aims to avoid mixing EPU triples between unrelated articles within an issue (\ie, it controls for false positives).\footnote{We use $\tau = 125$ tokens, the median number of tokens in Radio-Canada articles. For Radio-Canada, we set $\tau = \infty$ as this source is available at the article level. We also test alternative setups, and find that results are similar for values ranging from $\tau=50$ to $\tau = 1,000$. Lower values are too restrictive, and larger values result in triples being observed in unrelated news articles, leading to a noisier index.} From this subset, to avoid multiple counting, we count the number of unique 
triples:  $\textit{Tri}_{d,t,s}$ denotes the count in document~$d$ published in month~$t$ by source~$s$.\footnote{A unique triple is defined
as a set of three unique token positions. For instance, consider two triples with positions \{1, 4, 8\} and \{1, 7, 10\}. They count as a unique triple since position 1 appears in both. In other words, a token position can only be used once.} Then, following \citet{BakerEtAl2016}, we aggregate and scale our counts by source each month to get a monthly source-specific frequency-count measure:
\begin{equation}
F_{t,s} = \frac{\sum_{d = 1}^{D_{t,s}} \textit{Tri}_{d,t,s}}{\sum_{d = 1}^{D_{t,s}} \textit{Tok}_{d,t,s}} \,,
\end{equation}
where $\textit{Tok}_{d,t,s}$ denotes the total number of tokens for document~$d$ published in month~$t$ by source~$s$, and $D_{t,s}$ denotes the total number of documents published in month~$t$ by source~$s$.

\paragraph{Source aggregation} 

In \citet{BakerEtAl2016}, the source-specific frequency-count variables $F_{t,s}$ are first standardized by their standard deviation computed over a fixed and common time period. They are then aggregated to build the EPU index.\footnote{The fixed-time window ranges from 1985 to 2009 for the American EPU, while the Canadian EPU relies on data before 2011.} The standardization aims to account for differences in the variability of sources' reporting. However, the availability of our sources differs across time periods. Hence, the variability but also the level of $F_{t,s}$ are period-specific, making the traditional aggregation biased.\footnote{For instance, consider two sources whose EPU count frequency levels are different (one high and one low). If the low-count source disappears from the sample, the aggregate EPU will increase, despite the absence of an EPU event. We refer to the Internet appendix for illustrations.} To tackle this, we propose a dynamic approach. In month~$t$, we first scale $F_{t,s}$ by its $m$-past months (rolling-window) standard deviation to ensure similar variation across sources. Second, we divide the scaled variable by its $m$-past months average to ensure similar levels across sources.\footnote{We set $m = 36$ months (\ie, three-year rolling windows). An alternative setup with $m=60$ leads to similar graphical results with a correlation of 0.94, and nonsignificant different nowcasting performance; see the Internet appendix.} Then, we aggregate the source-specific scaled and normalized frequency-count measures, $\textit{EPU}_{t,s}$, to construct the EPU index in month~$t$:
\begin{equation}
\textit{EPU}_t = \sum_{s = 1}^{S_t}\frac{\textit{EPU}_{t,s}}{S_t} \,,
\end{equation}
where $S_t$ denotes available sources  in month $t$.

Our EPU construction ensures the index in month $t$ is only based on data available up to month $t$ (\ie, it is not forward-looking biased), which can be critical for practical applications such as forecasting. Note that the length of the rolling-window $m$ makes the interpretation of the EPU relative to its past $m$-month values.\footnote{We set $m=36$ months in our empirical applications. Results with $m=60$ are qualitatively similar and are available in the Internet appendix.} We believe it is a reasonable approach compared to selecting a single fixed reference period, especially for such a long time frame. The dynamic normalization accounts for a possible evolution in the media's writing style and type of news coverage.\footnote{In the Internet appendix, we show that the dynamic approach better identifies historical events/crises and provides better nowcasting performance than a fixed-window approach as in \citet{BakerEtAl2016}.} 

\section{Validation of the  index}
\label{sec:empres}

\begin{figure}[H]
\centering
\includegraphics[width=\textwidth]{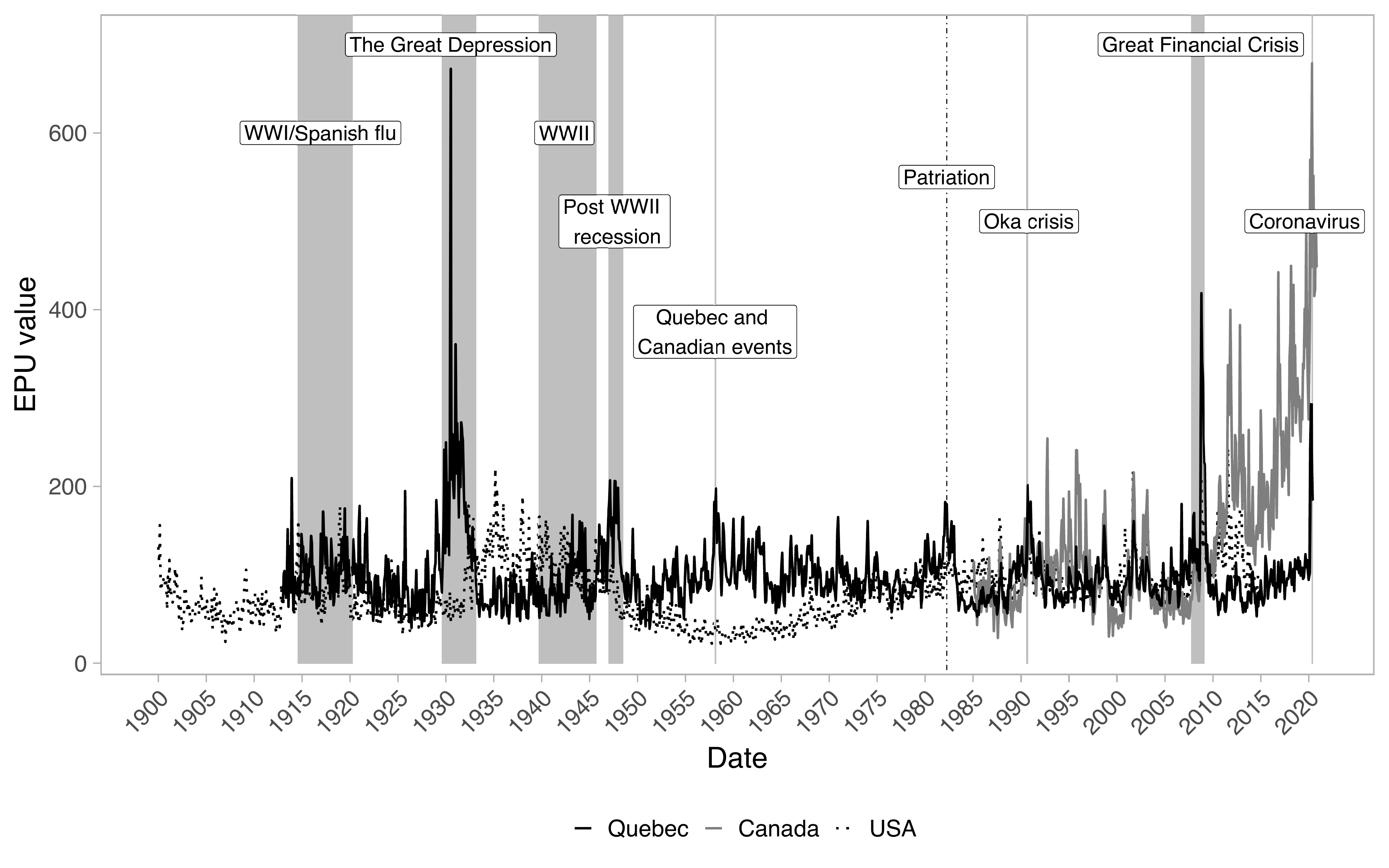}
\caption{\textbf{EPU indices and events}\\
This figure displays our EPU index (solid black line, from January 1913 to June 2020 for a total of 1,290 observations) together with the Canadian EPU index (solid gray line, from January 1985 to May 2020 for a total of 425 observations) and the historical American EPU index (dotted line, from January 1900 to October 2014, for a total of 1,378 observations).}
\label{fig:epu}
\end{figure}

\noindent 
In Figure~\ref{fig:epu}, we display the evolution of our EPU index (January 1913 to June 2020) together with the historical American EPU (January 1900 to October 2014) and the Canadian EPU (January 1985 to June 2020).\footnote{The American and Canadian EPU indices are available at \url{https://www.policyuncertainty.com}.} We see that our index spikes at major economic events. When looking over our 100-year time horizon, three peaks are particularly large: (i) the Great Depression, (ii) the 2008 financial crisis, and (iii) the COVID-19 pandemic. Moreover, the index tends to be higher post-war and during some other events such as the patriation of the Canadian constitution in 1982 (a national event) or the Oka crisis in 1990 (a Quebec-specific event). In comparison, the American index does not spike during the Great Depression or the post-war recession, raising questions about its accuracy.\footnote{According to \url{https://www.policyuncertainty.com}, the American historical index is still a work in progress.} The Canadian EPU seems to track the Quebec EPU until 2010, when we start to see a large discrepancy whereby the Canadian EPU increases with no economic rationale. This discrepancy could be attributed to differences in calculation methodologies and/or data sources used to build the indices. A change in the media landscape could also explain this upward trend in the Canadian EPU, justifying the need for a dynamic normalization approach.

\insertfloat{Figure~\ref{fig:epu}}

\section{Nowcasting macroeconomic variables}

\noindent 
We now investigate our EPU index's effectiveness in nowcasting five monthly macroeconomic variables: (i) log-changes in Canadian gross domestic product (GDP), (ii) log-changes in Canadian and Quebec consumer price indices (CPIs), and (iii) changes in Canadian and Quebec unemployment rates (UNEMPs).\footnote{Quebec GDP is not available at the monthly frequency.} We use the following specification:
\begin{equation}\label{eq:M0}
\mathcal{M}_0: \quad y_{t|t+1} = \beta_{0} +  \beta_{1} y_{t-1|t} +  \beta_{2} \textit{EPU}_{t|t} + \boldsymbol{\theta}' \mathbf{x}_{t-1|t} + \epsilon_{t}\,,
\end{equation}
where $y_{t|\bullet}$ is the macroeconomic variable of interest, $\mathbf{x}_{t-1|\bullet}$ is a vector of lagged explanatory variables, and $\epsilon_{t}$ is an error term. We consider a large set of Canadian and Quebec macroeconomic variables in $\mathbf{x}_{t-1|\bullet}$; see \citet{FortinGagnonEtAl2018}.\footnote{The full dataset is available at \url{https://www.stevanovic.uqam.ca/DS_LCMD.html}. In our nowcasting exercise, the number of available variables ranges from 130 to 140.} With our conditioning notation, we emphasize that all variables but the EPU are available with a time lag. We aim to evaluate out-of-sample nowcasting performance estimating the model on 60-month rolling windows starting in January 1985 and ending in June 2020.\footnote{We choose this time window to compare our EPU with the Canadian EPU.} This, however, results in a smaller sample size than the number of coefficients needed to be estimated, making the ordinary least squares framework unfeasible.

To deal with the high dimension, we proceed in two ways \citep[see][]{ArdiaEtAl2019}. First, we estimate the model with an elastic-net \citep{ZouEtAl2005}. Elastic-net is a penalized linear regression model that performs variable selection and shrinks regression coefficients towards zero, making the estimation of a model where the number of coefficients is larger than the number of observations feasible. The extent of the penalization is driven by hyper-parameters, which we select using the BIC-like criterion of \citet{ZouEtAl2007}. Second, we reduce the dimensionality of the explanatory variables using principal components. This transformation into a lower-dimensional space then allows estimating the model using ordinary least squares. We select the number of optimal principal components following \citet{BaiNg2002}.

Performance results are reported in Table~\ref{tab:perf} with our baseline model $\mathcal{M}_0$ together with a model $\mathcal{M}_1$ without EPU (\ie, $\beta_2 = 0$) and a model $\mathcal{M}_2$ for which the EPU is the Canadian EPU. First, for both the elastic-net and the principal components 
approaches, we find that our EPU-enhanced model has better results than the model using only macroeconomic variables ($\mathcal{M}_0$ vs. $\mathcal{M}_1$). This holds for the root-mean-square forecast error (RMSFE) and the mean absolute forecast error (MAFE) performance metrics. The improvement 
is statistically significant at the 5\% level for Canadian GDP and for Canadian and Quebec unemployment. Hence, our Quebec EPU helps to nowcast macroeconomic variables at the provincial and national levels. Next, we find that our EPU outperforms the nowcasting ability of the traditional Canadian EPU  ($\mathcal{M}_0$ vs. $\mathcal{M}_2$). For both the estimation methods or the performance metrics, the outperformance is significant at the 5\% level for Canadian GDP and unemployment rates for Quebec and Canada.\footnote{Additional analyses reported in the Internet appendix show that similar nowcasting results are obtained with alternative rolling-window lengths, while a static-window approach deteriorates the performance.}

\insertfloat{Table~\ref{tab:perf}}

\section*{A. Internet appendix}

\noindent
The Internet appendix contains additional information about the database, methodology and nowcasting performance of the models. The EPU index is available at \url{https://sentometrics-research.com}.


\newpage
\begin{table}[H]
\caption{\textbf{Nowcasting results}\\
The table reports the root-mean-square forecast error (RMSFE) and the mean absolute forecast error (MAFE) for the various models applied to five monthly macroeconomic variables: Canadian gross domestic product (GDP, log-changes), consumer price index (CPI, log-changes), and unemployment rate (UNEMP, changes); and Quebec CPI and UNEMP. We consider $\mathcal{M}_0$ in \eqref{eq:M0} and two alternatives: $\mathcal{M}_1$ is without EPU (\ie, $\beta=0$) and $\mathcal{M}_2$ is with the Canadian EPU instead of our EPU. Squared parentheses report the p-value of a Diebold-Marino test of outperformance of $\mathcal{M}_0$ against $\mathcal{M}_\bullet$ using the approach
described in \citet{ArdiaEtAl2019}. The out-of-sample performance window ranges 
from January 1990 to June 2020 for a total of 366 observations (the first rolling window used to estimate the 
models ranges from January 1985 to December 1989).}
\centering
\scalebox{0.95}{
\begin{tabular}{lcccccc}
\toprule
\multicolumn{7}{l}{Panel A: Regularization (elastic-net)}\\
&  \multicolumn{3}{c}{RMSFE (x100)} & \multicolumn{3}{c}{MAFE (x100)} \\
\cmidrule(l{5pt}r{5pt}){2-4}
\cmidrule(l{5pt}r{5pt}){5-7}
Variable  & $\mathcal{M}_0$ & $\mathcal{M}_1$  & $\mathcal{M}_2$ & $\mathcal{M}_0$ & $\mathcal{M}_1$  & $\mathcal{M}_2$\\
\midrule
GDP (CAN) & 1.041 &  1.415[0.000] & 1.429[0.002] &  0.341 &   0.383[0.002] &  0.384[0.096]\\
CPI (CAN) & 0.369 &  0.367[0.828] &  0.379[0.038] & 0.272&  0.272[0.613] &  0.282[0.043]\\
UNEMP (CAN) & 0.441 & 0.604[0.000] &  0.608[0.000]  & 0.177 &  0.193[0.001] &  0.194[0.000]\\
CPI (QC) & 0.425 &  0.425[0.585] &  0.440[0.016] &  0.299  &  0.302[0.415] &  0.316[0.075]\\
UNEMP (QC) & 0.989 &  1.254[0.000] &  1.264[0.000] &  0.334 &   0.353[0.001] & 0.358[0.000]\\[0.3cm]
\multicolumn{7}{l}{Panel B: Principal components}\\
&  \multicolumn{3}{c}{RMSFE (x100)} & \multicolumn{3}{c}{MAFE (x100)} \\
\cmidrule(l{5pt}r{5pt}){2-4}
\cmidrule(l{5pt}r{5pt}){5-7}
Variable  & $\mathcal{M}_0$ & $\mathcal{M}_1$  & $\mathcal{M}_2$ & $\mathcal{M}_0$ & $\mathcal{M}_1$  & $\mathcal{M}_2$\\
\midrule
GDP (CAN) & 1.139 &  1.506[0.000] &  1.479[0.000] & 0.349  &  0.382[0.000] & 0.380[0.000]\\
CPI (CAN) & 0.371 &  0.377[0.101] &  0.382[0.026] & 0.273  &  0.277[0.205] & 0.280[0.030]\\
UNEMP (CAN) & 0.435 & 0.591[0.001] &  0.584[0.000] & 0.171 & 0.186[0.000] &  0.185[0.000]\\
CPI (QC) & 0.431 &  0.433[0.364] &  0.439[0.143] & 0.306 & 0.308[0.354] & 0.316[0.062]\\
UNEMP (QC) & 1.017 &  1.250[0.000] &   1.241[0.000]  & 0.336 & 0.342[0.003] & 0.347[0.000]\\
\bottomrule
\end{tabular}}
\label{tab:perf}
\end{table}

\newpage

\end{document}